\documentclass[aps,twocolumn,raggedbottom,prl,showpacs,nobalancelastpage,amssymb,groupedaddress]{revtex4}
\usepackage[english]{babel}
\usepackage{graphicx}
\usepackage{amssymb}
\usepackage{amsmath}
\usepackage{amscd}
\usepackage{eucal}
\usepackage{color}
\usepackage{bm}
\usepackage{upgreek}
\usepackage{subfigure}
\usepackage{wasysym}

\def\be{\begin{equation}}
\def\ee{\end{equation}}
\def\bea{\begin{eqnarray}}
\def\eea{\end{eqnarray}}
\def\bsub{\begin{subequations}}
\def\esub{\end{subequations}}

\newcommand{\dd}[2]{\frac{\partial #1}{\partial #2}}

\begin{document}

\title{Interplay between non equilibrium and equilibrium spin torque using~synthetic~ferrimagnets}

\newcommand{\spsms}{SPSMS, UMR-E 9001 CEA / UJF-Grenoble 1, INAC, Grenoble, F-38054, France}
\newcommand{\karl}{Physikalisches Institut,  Universit\"at  Karlsruhe (TH), Wolfgang-Gaede Strasse 1, 76131 Karlsruhe, Germany}

\date{\today}

\author{Christian Klein}
\affiliation{\spsms,\karl}
\author{Cyril Petitjean}
\affiliation{\spsms}
\author{Xavier Waintal}
\affiliation{\spsms}
\date{\today}

\begin{abstract}
We discuss the current induced magnetization dynamics of spin valves  $ {\rm F_0|N|SyF}$  where the free layer is a synthetic ferrimagnet ${\rm SyF}$ made of two ferromagnetic  layers ${\rm F_1}$ and ${\rm F_2}$ coupled by RKKY exchange coupling. In the interesting situation where the magnetic moment of the outer layer ${\rm F_2}$ dominates the magnetization of the ferrimagnet, we find that the sign of the effective spin torque exerted on the free middle layer ${\rm F_1}$ is controlled by the strength of the RKKY coupling: for weak coupling one recovers the usual situation where spin torque tends to, say,  anti-align  the magnetization of ${\rm F_1}$ with respect to the pinned layer ${\rm F_0}$. However for large coupling the situation is reversed and the spin torque tends to  align ${\rm F_1}$ with respect to ${\rm F_0}$. Careful numerical simulations in the intermediate coupling regime reveal that the competition between these two incompatible limits leads generically to spin torque oscillator (STO) behavior. The STO is found in the absence of magnetic field, with very significant amplitude of oscillations and
frequencies up to 50 GHz or higher.  
\end{abstract}


\maketitle

Since the first successful experimental manipulation of magnetic configurations  by  spin-polarized currents~\cite{Tsoi:1998,Myers:1999}, the interest in spintronics devices entirely controlled by electrical currents  has rapidly increased~\cite{Katine:2000,Tsoi:2000,Kiselev:2003,Rippard:2004,Krivorotov:2005,Boulle:2007, Pribiag:2007,Houssameddine:2007,Rippard:2010,Madami:2011}. The key concept behind theses experiments is the notion of spin transfer torque (STT) predicted by  Slonczewski~\cite{Slonczewski:1996} and Berger~\cite{Berger:1996}: a current gets spin polarized by a first (pinned) magnetic layer and then transfers some of its polarization to a second free layer, hence exerting a torque on its magnetization.   As a result the magnetization can switch to a different static configuration~\cite{Myers:1999,Katine:2000} or  even to a dynamical stationary regime where the magnetization of the free layers
shows a sustained precession~\cite{Tsoi:2000,Kiselev:2003,Rippard:2004,Krivorotov:2005}.  This dynamical regime, known as the spin torque oscillation (STO), allows 
to convert a dc voltage into a microwaves signal at several GHz and has a real potential for applications (nanoscale microwave source or detector, high frequencies possibly
above CMOS technology, narrow bandÉ). There are still technological issues with STO based devices, however, such as very limited output power and the need of rather strong external fields.
Indeed, the main mechanism for producing STO is based on a detailed tuning of the angular dependance of the spin torque~\cite{Kiselev:2003}  and requires the use of an external magnetic field.  Recently, different mechanisms have emerged to enable STO behavior in the absence of external field: the first approach~\cite{Pribiag:2007} relies on the precession of a non uniform magnetic structure, a magnetic vortex, that can be excited at sub GHZ frequencies;  a second approach~\cite{Boulle:2007,Rychkov:2009} uses  a strongly asymmetric spin valve to increase the anharmonicity of the angular dependance of the spin torque,  leading to the so-called  "wavy" behavior of the STT~\cite{Manschot:2004, Barnas:2005, Gmitra:2006,Rychkov:2009}. Other approaches use an orthogonal polarizer~\cite{Houssameddine:2007} or a perpendicular free layer~\cite{Consolo:2007,Rippard:2010}.

In this  letter we propose an entire new mechanism which naturally induces strong STO behavior, even in the absence of magnetic field, using a synthetic ferrimagnet (${\rm SyF}$) as the free layer of the spin valve. The ${\rm SyF}$ is  made of two ferromagnetic layers coupled through antiferromagnetic RKKY exchange interaction~\cite{Parkin:1990, Parkin:1991e,Bruno:1992}.  Here, we predict that by simply tuning the effective strength of  the RKKY exchange coupling (through the different thicknesses of the layers for instance), 
the {\it sign } of the effective STT can be changed. At intermediate effective coupling, the STT generically induces strong STO behavior irrespectively of the presence of applied magnetic field.

In the following we first provide a simple physical picture that captures the essential features of our setup. Our main prediction uses very general conservation arguments and should therefore be quite robust and independent of the details of the model. In a second step, we provide more quantitative arguments to describe the strong coupling regime. Last, we present a detailed numerical study of the phase diagram of our model. We find strong STO behavior at rather large frequencies in a wide portion of the phase diagram hinting that ${\rm SyF}$ based STOs could be good candidates for application purposes.

\begin{figure}[h]
\begin{center}
\includegraphics[angle=0,width=1\linewidth]{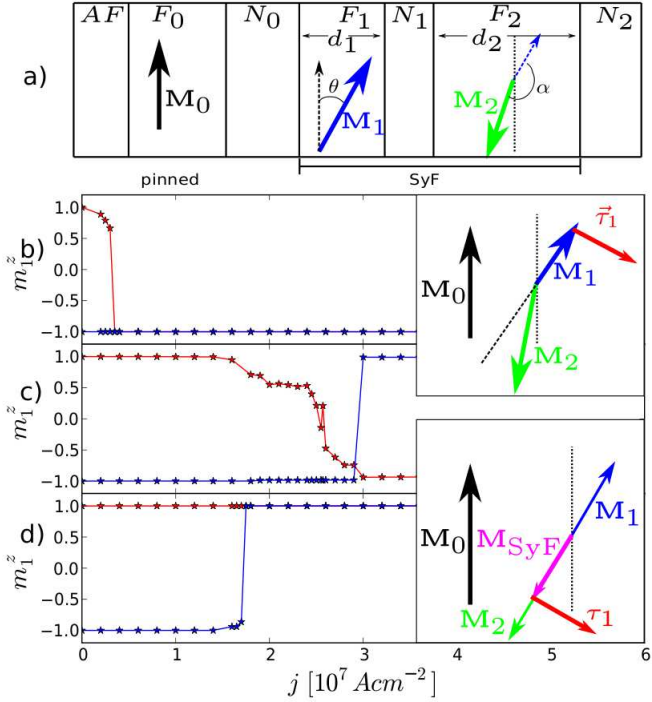}
\caption{
Panel (a) is a cartoon of the magnetic multilayer stack, in which   ${\bf M}_i$ is the  the magnetization of the  layer ${\rm F_i}$.   The angle between ${\bf M}_1$ and ${\bf M}_0$ (${\bf M}_2$) is respectively $\theta$ ($\alpha$).  Panels  (b,c,d)  show the stationary magnetization $m_{1}^z$ along ${\bf M}_0$ as  a function of  the current density $j$ in  ${\rm A.cm^{-2}}$, for a stack with $d_1=2.~{\rm nm}$,  $d_2=4\ d_1$ and three different coupling strength J; (b)  weak coupling limit ($J=0$), (c) Intermediate  coupling  limit ($J=-10^{-3}~{\rm J.m^{-2}}$) (d) Strong  coupling  limit ($J=-6.\ 10^{-3}~{\rm J.m^{-2}}$)
Upper (Lower)  inset : Sketch of the torque $\pmb{\tau}_{1}$ action on ${\bf M}_1$ (${\bf M}_{\rm SyF}$) for the weak (strong) coupling regime.
}
\label{fig:stack}
\end{center}
\end{figure}

{\it Physical mechanism. }
We consider a nanopillar spin valve ${\rm F_0| N_0| SyF}$ made of a pinned ferromagnetic layer  ${\rm F_0}$ and a free synthetic ferrimagnetic layer ${\rm SyF= F_1 | N_1 | F_2}$ (where ${\rm N_0}$ and ${\rm N_1}$ are normal spacers), see Fig.~\ref{fig:stack}a for a cartoon of the full stack. Upon injecting a current density $j$ through the nanopillar, a spin torque $\pmb{\tau}$ is exerted on the magnetic moments of the layers~\cite{caveat0}. This torque has been extensively discussed in the literature and can be strong enough
to destabilize the initial magnetic configuration. On the other hand, the magnetization $\mathbf{M}_1$ and $\mathbf{M}_2$ of  ${\rm F_1}$ and  ${\rm F_2}$ are coupled by 
an oscillatory exchange (RKKY) interaction $J$ which is nothing but the spin torque present in the pillar {\it at equilibrium}~\cite{Slonczewski:1989,Waintal:2002}. The coupling $J$ is reminiscent of
Friedel oscillations and typically behaves as $J\sim (\cos 2k_Fd_N)/d_N$ where $k_F$ is the Fermi momentum and $d_N$ the width of the spacer $N_1$. $J$ can be tuned from negative (antiferromagnetic coupling) to positive  (ferromagnetic coupling), hence the name "oscillatory". Here we focus on negative values of $J$ which stabilize an anti-aligned configuration of $\mathbf{M}_1$ and $\mathbf{M}_2$. The goal of this letter is
to discuss  new phenomena that arise from the competition between the equilibrium and the non equilibrium torques.

In order to gain a qualitative understanding of the underlying physics, let us discuss two extreme limits.  We start from a configuration where $\mathbf{M}_1$ is aligned
with $\mathbf{M}_0$ while $\mathbf{M}_2$ is anti-aligned, and inject electrons from the right.  Upon crossing the pinned layer ${\rm F_0}$, the electrons get a polarization 
anti-parallel to $\mathbf{M}_0$. Denoting $\mathbf{J}_{i}$ the spin current (per unit surface) flowing in the normal layer ${\rm  N_{i}}$, we have 
$\mathbf{J}_{0}=\hbar pj \mathbf{m}_0/(2e)$ with $\mathbf{m}_0={\bf M}_0/|{\bf M}_0|$ and $p$ the polarization of the current ($0\leq p \leq 1$). Let us focus on the weak coupling limit where $J\rightarrow 0$.
When there is a finite angle $\theta$ between  $\mathbf{M}_0$ and $\mathbf{M}_1$, the spin current on the left $\mathbf{J}_{0}$ and right $\mathbf{J}_{1}$ of ${\rm F_1}$ become different and a torque $\pmb{\tau}_1=\mathbf{J}_{0}-\mathbf{J}_{1}$ is exerted on 
the magnetization of ${\rm F_1}$: conservation of angular momentum implies that whatever spin  lost by the conducting electrons is gained by magnetic degrees of freedom~\cite{Slonczewski:1996}.   This conservation of angular momentum reads,
\be
\frac{\partial}{\partial t} \left[ \frac{d_1\mathbf{M}_1}{\gamma_1} \right] =\pmb{\tau}_1 +\dots
\label{torque1}
\ee
where $d_1$ is the width of ${\rm F_1}$, $\gamma_1$ its gyromagnetic ratio and the $\dots$ indicate other contribution to the dynamics (magnetic anisotropy, dampingÉ) to be discussed later (upper inset of Fig.~\ref{fig:stack}).  Ignoring (momentarily \cite{caveat}) the role of ${\rm F_2}$, we arrive at $\partial\theta/\partial t = \gamma_1\hbar p j \sin \theta/(2e d_1 \vert M_1\vert)$ and recover the usual phenomenology
of spin torque: the torque tends to stabilize the configuration where $\mathbf{M}_1$ is anti-parallel  to $\mathbf{M}_0$ (or destabilize it if one reverses the sign of the current). 

Let us now discuss how this picture is modified when one considers the opposite strong coupling $J\rightarrow -\infty$ limit. Upon switching on $J$, 
the exchange energy (per unit surface of the pillar) $E_{J}  = -J\,  (\mathbf{m}_1\cdot \mathbf{m}_2 )$ induces a field like term in the RHS of Eq.(\ref{torque1})
of the form   $+J \mathbf{m}_2 \times \mathbf{m}_1$. One needs also to consider the corresponding equation for the dynamics of $\mathbf{M}_2$ (see below Eq.(\ref{eqn:mdot_all})
for the full model) and one obtains a potentially rich coupled dynamics for $\mathbf{M}_1$ and $\mathbf{M}_2$. The situation simplifies in the limit where
the exchange energy $J\rightarrow -\infty$ dominates: the relative configuration of $\mathbf{M}_1$ and $\mathbf{M}_2$ becomes frozen in the anti-aligned position. Conservation of angular momentum now leads to a spin torque on the total magnetic moment:
\be
\frac{\partial}{\partial t} \left[ \frac{d_1\mathbf{M}_1}{\gamma_1} +\frac{d_2\mathbf{M}_2}{\gamma_2} \right] =\pmb{\tau}_1 +\dots
\label{torque2}
\ee
where we have used the fact that $\pmb{\tau}_2=\mathbf{J}_{1}-\mathbf{J}_{2}$ vanishes when $\mathbf{M}_1$ and $ \mathbf{M}_2$ are collinear.
Note that the exchange field, which conserves the total angular momentum, redistributes the torque $\pmb{\tau}_1$ on $\mathbf{M}_1$ and $ \mathbf{M}_2$ so that
they remain anti-aligned. 

Eq.(\ref{torque2}) allows two very different regimes: if $d_1 \vert \mathbf{M}_1 \vert /\gamma_1 > d_2 \vert  \mathbf{M}_2\vert /\gamma_2$ then the effective magnetization
direction $\mathbf{M}_{\rm SyF}$ of the SyF layer points along $\mathbf{M}_1$ and the torque  is very similar to the weak coupling regime. This case has attracted some attention recently both experimentally by Smith et al.~\cite{Smith:2008} and theoretically by Balaz et al.~\cite{Balaz:2011}. 
Here, we focus on the other limit $d_1 \vert \mathbf{M}_1 \vert /\gamma_1 < d_2 \vert  \mathbf{M}_2\vert /\gamma_2$ where $\mathbf{M}_{\rm SyF}$ points in the 
{\it opposite } direction as $\mathbf{M}_1$ (lower inset of Fig.~\ref{fig:stack}). In this limit, the torque tends to stabilize the configuration where $\mathbf{M}_{\rm SyF}$ is
anti-aligned with $\mathbf{M}_0$ hence where $\mathbf{M}_0$ and $\mathbf{M}_1$ are aligned: this is the opposite situation from the weak coupling $J\rightarrow 0$ limit.

To summarize, when going from $J\rightarrow 0$ to $J\rightarrow -\infty$, without changing the sign of the current, the torque tends to favor two different stationary situations.
In the rest of this letter, we study this crossover in more details. In particular, we find that the intermediate coupling regime reveals interesting, STO like, dynamical behavior.

{\it Model.} 
We now turn to the modelisation of our system. The full stack we are considering has the form (see Fig.~\ref{fig:stack}a): ${\rm AF | F_0 | N_0 | F_1 | N_1 | F_2| N_2 }$,  where ${\rm AF}$ is an antiferromagnetic layer (typically IrMn) that pins the magnetization of ${\rm F_0}$.
We set our  reference  spin axis $\mathbf{e}_z$ parallel to $\mathbf{M}_0$.
Magnetization dynamics is described by two coupled Landau Lifshiz Gilbert (LLG) equations that read  (in SI units),
\bsub\label{eqn:mdot_all}
\bea
\dd{\mathbf{m}_1}{t}
	&=& \gamma_1  \mathbf{B}_1 \times \mathbf{m}_1
	+ \alpha_1 \mathbf{m}_1 \times \dd{\mathbf{m}_1}{t}	+ \frac{\gamma_1}{\vert \mathbf{M}_1 \vert d_1}\pmb{\tau}_1\quad
	\label{eqn:mdot_1}
\\
\dd{\mathbf{m}_2}{t}
	&=& \gamma_2  \mathbf{B}_2 \times \mathbf{m}_2
	+ \alpha_2 \mathbf{m}_2 \times \dd{\mathbf{m}_2}{t}
	+   \frac{\gamma_2}{\vert \mathbf{M}_2 \vert d_2} \pmb{\tau}_2,\quad
	\label{eqn:mdot_2}
\eea
\esub
where $\mathbf{m}_i = \mathbf{M}_i/ \vert \mathbf{M}_i \vert$ is the unit vector describing the magnetization orientation of the  layer $\bf{ F_i}$
and $\alpha_i$ is the damping factor of the corresponding magnetic material.   The effective field $\mathbf{B}_i$ seen by the magnetization in the layer ($i=(1,2)$)
is 
\be\label{eqn:magfield}
\mathbf{B}_i = \frac{K^i_u}{\vert \mathbf{M}_i \vert}  \left[\mathbf{m}_i \cdot \mathbf{e}_z\right] \mathbf{e}_z +  \frac{J}{\vert \mathbf{M}_i \vert d_i} \mathbf{m}_{\bar i}\,,
\ee
where $\bar i=3-i$ corresponds to the index of the other layer. The effective field includes an uniaxial anisotropy $K^i_u$ field and the RKKY exchange field.    
In the physical regimes studied below, RRKY coupling is the dominating energy so that our results are rather insensitive to the presence of less important terms
(for instance dipolar coupling between layers that we do not take into account).
The  spin torque $\pmb{\tau}_{i}$ is calculated  in the framework  of CRMT (Continuous Random Matrix Theory)~\cite{Rychkov:2009, Borlenghi:2011}.
 CRMT is a  semiclassical approach that generalizes Valet-Fert theory to non collinear situations. For discrete systems, It is formally equivalent to the generalized circuit theory~\cite{Bauer:2003b} and is well adapted to the treatment of metallic magnetic multilayers. 

{\it Strong coupling limit.} 
When the RKKY coupling is large, one can derive an effective LLG equation for the
dynamics of the SyF. It  reads,
\be
\label{eqn:m1_eff}
\dd{\mathbf{m}_1}{t} = \gamma_{\rm eff}  \mathbf{B}_{\rm eff}  \times \mathbf{m}_1
	     + \alpha_{\rm eff}  \mathbf{m}_1 \times \dd{\mathbf{m}_1}{t}	+ \frac{\gamma_{\rm eff} }{\vert \mathbf{M}_1 \vert d_1}\pmb{\tau}_1, \quad
\ee
where the various effective parameters get renormalized as follows:
$ \gamma_{\rm eff}  =\gamma_{1} /(1-\delta)$,  $\alpha_{\rm eff}  = ( \alpha_1 +\delta \alpha_2)/(1-\delta)$,  $K^{\rm eff}_u =K^1_u + K^2_u d_2/d_1$, $
\mathbf{B}_{\rm eff} =  ( K^{\rm eff}_u/ \vert \mathbf{M}_1\vert) [\mathbf{m}_1\cdot \mathbf{e}_z ]   \mathbf{e}_z$
and the renormalization parameter $\delta$ has been defined as,
 \be
 \delta = \frac{\gamma_1\vert \mathbf{M}_2 \vert d_2}{\gamma_2\vert \mathbf{M}_1 \vert d_1}
 \ee
(Eq.(\ref{eqn:m1_eff}) is obtained by calculating $\partial/\partial t [\mathbf{m}_1 +\delta \mathbf{m}_2]$ which cancels the RKKY contribution, and then setting $\mathbf{m}_2= -\mathbf{m}_1$ and $\pmb{\tau}_2= \mathbf{0}$).
From this equation we can clearly see the inversion of  the sign of the effective torque $\gamma_{\rm eff} \pmb{\tau}_1$ discussed previously when $\delta > 1$.

  \begin{figure}[h]
\begin{center}
\rotatebox{0}{\resizebox{8cm}{!}{\includegraphics{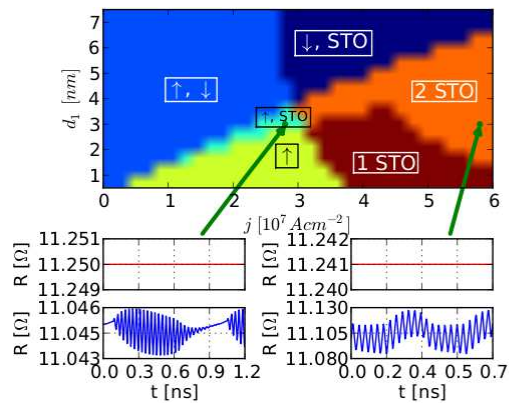}}}
\caption {Upper:  phase diagram of the system as a function of the current density $j$ and the thickness $d_1$ for a fixed $\delta = 4$ and  $J=-5.~10^{-3} ~{\rm J.m^{-2}}$. The various symbols indicate the observed stationary states in the corresponding region (defined by different colors): $\uparrow$ ($\downarrow$) stands for $m_1^z=1$ ($m_1^z=-1$), while $STO$  corresponds to the presence a precessional state. In the 2 $STO$ region, two different STO states can be observed depending on the hysteresis history. Lower:  Resistance $R$ as function of time $t$ for a  pillar surface of $1000~{\rm nm}^2$ and $d_1=3 {\rm nm}$. Upper (lower) plots correspond to initial conditions in the up (down) configuration.
Left panels:  $j=2.8\ 10^7 {\rm A.cm^{-2}}$. Right panels: $j=5.8\ 10^7 {\rm A.cm^{-2}}$. The amplitude of the oscillation with the up initial condition is small and invisible on the scale of the graphics }
\label{fig:phases}
\end{center}
\end{figure}
{\it Numerical simulations.} 
Let us now go back to the full model Eq.(\ref{eqn:mdot_all}) and study it numerically.
We focus on a stack defined as ${\rm IrMn_{\rm 5}\vert Py_{\rm 5}\vert  Cu_{\rm 10}\vert  Co_{\rm d_1}\vert  Ru_{\rm0.3}\vert Co_{\rm d_2}\vert  Cu_{\rm 10 } }$ where the index indicate the
widths of the layers in ${\rm nm}$. We have used  $K_u^{\rm Co} = 8.~10^4~{\rm J.m^{-3}}$,  $M_{\rm Co}=1.42~10^6~{\rm A.m^{-1}}$  and the transport parameters (spin resolved bulk resistivities, interface resistivities, spin flip lengthsÉ) have been taken from the database established in MSU and CNRS/Thales laboratory~\cite{Jaffres}.  The coupling constant $J$ can be tuned by changing the $Ru$ thickness with typical values $J^{\rm Ru}\approx -5.\ 10^{-3}~{\rm J.m^{-2}}$~\cite{Parkin:1991e}. The torque and resistances
have been calculated using CRMT and the corresponding parametrization (as a function of the angles $\theta$ and $\alpha$, see Fig.\ref{fig:stack}a) incorporated into our
numerical LLG integrator.

A first set of curve is presented in Fig.~\ref{fig:stack} b,c and d where the stationary magnetization $m_1^z$ along the $z$ axis is plotted against the current density $j$ for three
values of the coupling constant: small (b), intermediate (c) and large coupling (d). For small coupling (b)  , we recover the usual behavior of current induced magnetic reversal in spin valves: at zero current the system has two stable configurations where $\mathbf{m}_1$ is aligned ($m_1^z=1$) or anti aligned ($m_1^z=-1$) 
with $\mathbf{m}_0$. Upon injecting a strong positive current, one "pushes" toward the anti-aligned configuration which becomes stable above a critical value of the current 
density $j$ ($\rm 0.5\ 10^{7} A.cm^{-2}$ in this example). At large coupling (d), this hysteretic curve is reversed, in agreement with the arguments developed above. 
A very interesting regime is found at intermediate couplings (c) where one observes a rather large window where the stationary value of $m_1^z$ corresponds to a finite angle between $\mathbf{m}_0$ and $\mathbf{m}_1$, i.e. to a dynamical state where a sustained precession of $\mathbf{m}_1$ is present (with frequencies around 15 GHz in this example).

The coupling $J$ can be varied by tuning the width $d_{\rm N}$ of the normal (Ru) spacer. However, this is not very easy to control in practice, as the oscillatory character of
the RKKY interaction make this variation non monotonic. Alternatively, one can explore the phase diagram of our stack by varying the thickness $d_1$ for a fixed ratio $\delta=d_2/d_1$, hence varying the relative importance of bulk versus surface terms in the LLG equation. Fig. \ref{fig:phases}, shows the resulting phase diagram in the $(j,d_1)$ plane for 
$\delta=4$. The different symbols indicate the possible stationary states in the various regions of the phase diagram (defined by the various colors).  The presence of two different symbols (most regions) indicate that depending on the hysteresis history, one or the other state is observed. In particular, in the 2 STO region, two different STO states can be observed depending, for instance, on the initial condition $m_1^z(t=0)=\pm1$. These two different STO states merge upon entering the 1 STO region. 
We have investigated other stacks with different material parameters (not shown) and found very similar phase diagrams, including the presence of high frequency STO phases, as long as one remains in the $\delta>1$ regime. We observe two kind of STO behaviors, as evident from the Resistance versus time $R(t)$ traces plotted in Fig. \ref{fig:phases}. When the amplitude of the oscillating signal is small (lower left of Fig. \ref{fig:phases}), we observe some beating behaviors with one well defined frequency and a second much smaller frequency less well defined (corresponding to the precession of the two layers respectively). When these two frequencies become closer, the time dependent signal gets bigger and $R(t)$ becomes much more sinusoidal indicating phase locking between the precessional dynamics of the two magnetizations $\mathbf{m}_1$ and $\mathbf{m}_2$. 

\begin{figure}[h]
\begin{center}
\rotatebox{0}{\resizebox{7cm}{!}{\includegraphics{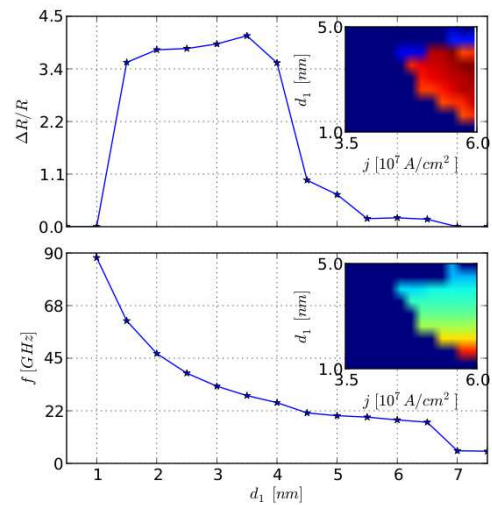}}}
\caption{Upper plot: amplitude (in $\permil$)  of the time dependent signal of the resistance as a function of $d_1$ for fixed $j=6.\ 10^7{\rm A. cm}^{-2}$.
Lower plot: idem for the main oscillating frequency of the resistance. Insets: corresponding colorplots in the $(j,d_1)$ plane. 
$\delta =4$ and $J=-5.\ 10^{-3}~{\rm J.m^{-2}}$.}
\label{fig:fft}
\end{center}
\end{figure}

In the last Fig.\ref{fig:fft}, we focus on the region of the phase diagram (lower right corner of Fig.\ref{fig:phases}) 
where the amplitude of the STO signal is the highest. The upper curves show the amplitude of the time
dependent signal while the lower curves show the corresponding frequency. We find that this large angle STO phase, which is stable in the absence of magnetic field, leads to very high frequencies up to several tens of GHz while sustaining  high precession angles (of the order of $\pi/6$ ). The observed high frequencies is tightly linked with the high value of the RRKY coupling which can be up to two orders of magnitude larger than typical anisotropy or Zeeman energies. Even if the relative angle between the two magnetizations of the synthetic ferrimagnet remain close to $\pi$, this gives rise to frequencies of several tens of GHz.

To conclude, we have shown that the interplay between equilibrium and non equilibrium torque in synthetic ferrimagnets based spin valves can lead to
interesting physics and potentially a new route to high frequency STOs.

 \begin{acknowledgments}
   We thank T. Rasing, W. Wulfhekel for very useful discussions.
   This work was supported by , CEA NanoSim program,   CEA Eurotalent and EC Contract Macalo.
    \end{acknowledgments}


\begin{thebibliography}{30}
\bibitem[{cav({\natexlab{b}})}]{caveat}
\bibinfo{note}{In the case considered here, the effect of $\pmb{\tau}_2$ will
  only slightly increase the predicted effect.}
\end{thebibliography}
\end{document}